# Electron cyclotron emission at the fundamental harmonic in GDT magnetic mirror

A. G. Shalashov[1,2, a)], A. L. Solomakhin[1], E. D. Gospodchikov[1,2],
L. V. Lubyako[1,2], D. V. Yakovlev[1] and P. A. Bagryansky[1]

[1]Budker Institute of Nuclear Physics (BINP), Siberian Branch of Russian Academy of Sciences, Lavrentieva ave. 11, 630090, Novosibirsk, Russia

[2]Institute of Applied Physics of Russian Academy of Sciences, Ulyanova str. 46, 603950 Nizhny Novgorod, Russia

**Abstract**

New electron cyclotron emission (ECE) diagnostics has been installed to facilitate the successful experiment on electron cyclotron plasma heating (ECRH) experiment at the large open magnetic trap GDT in Budker Institute. The particularities of ECE in the vicinity of the ECRH frequency were studied experimentally for a broad range of discharge scenarios. Measured thermal emission has partly validated the existing physical conceptions about the microwave plasma heating in the machine. Besides expected emission of thermal electrons, a clearly resolved non-thermal ECE was observed what unambiguously confirmed the presence of suprathermal electrons driven by high-power microwave heating.

a)Author to whom correspondence should be addressed. Electronic mail: ags@appl.sci-nnov.ru

## I. INTRODUCTION

The absorption of electromagnetic waves under the electron cyclotron resonance (ECR) conditions is widely used for heating of high-temperature plasmas in tokamaks and stellarators. However, for many years the use of this method in open magnetic configurations has been limited either to plasma heating in relatively compact laboratory installations [1], or to MHD stabilization and generation of fast electron population in a low-density plasma [2-5]. The exception was the TMX-U experiment at Lawrence Livermore National Laboratory where electron temperatures up to 0.28 keV were obtained with ECR heating (ECRH), but these studies were terminated soon after that [6]. Recently, efficient ECRH of a relatively dense plasma has been demonstrated in the large-scale axially symmetric mirror device GDT (gas-dynamic trap) in the Budker Institute of Nuclear Physics [7-11]. In these experiments, a combined plasma heating by neutral deuterium beams (total power up to 5 MW) and microwave radiation (700 kW) results in the record for open traps bulk electron temperatures up to 1 keV at density about $10^{13}$ cm$^{-3}$, thus demonstrating plasma parameters comparable to those in toroidal systems [10]. The main scientific goal of GDT is to validate the concept of a high-power source of fusion neutrons based on a magnetic mirror device [12]. Neutrons are born as a result of thermonuclear reactions between energetic deuterium ions driven by neutral beam injection (NBI); the energy confinement time of fast ions is determined by the electron-ion Coulomb collisions, $\tau_h \propto T_e^{3/2}$. For this reason, the electron temperature is the main factor limiting the confinement time of fast ions and, thus, the power efficiency of a beam-driven fusion reactor [13]. Success in ECRH experiments at GDT is convincing enough to consider the prospects of simple axially symmetric open magnetic traps as a high-power neutron source for a number of fusion applications [14], including material testing, a hybrid fusion-fission reactor and nuclear waste processing [12,15-18].

The ECRH system is operating at the fundamental harmonic of the extraordinary (X) mode. Implementing effective microwave heating of dense plasma at GDT requires a rather unusual scheme based on radiation trapping by a non-uniform plasma column [19, 20]; see also Ref. 21 in which the same scheme is studied using a more advanced theoretical model. This entirely new ECRH scenario substantially depends on a magnetic configuration and plasma inhomogeneity. Although successful ECRH experiments evidence for the reliability of such heating mechanism, a new electron cyclotron emission (ECE) diagnostics has been recently installed to improve understanding of complex wave physics related to ECRH in GDT. The original idea was to exploit the reciprocity principle and Kirchhoff's law that link the absorption and the spontaneous plasma emission at thermal equilibrium [22]. So, conditions favorable for the ECR plasma heating should manifest themselves as an increased ECE level of thermal electrons along the same ray. According to this idea, the ECE diagnostics operates near ECRH frequency (54.5 GHz) in the geometry reversed to the ECRH one. With this aim in mind, we connect the ECE receiver to a high-power transmission line instead of one of two available gyrotrons. Preliminary results of these experiments have been briefly reported in Ref. 23, while a full overview is presented below.

The paper is organized as follows: Sec. II presents general information about the GDT setup, details of the ECE hardware and experimental conditions for the ECE measurements; Sec. III and IV present the experimental data on cyclotron emission of thermal and suprathermal electrons, respectively, and a basic interpretation of the measurements; Sec. V summarizes the results.

## II. EXPERIMENTAL CONDITIONS

*A. ECRH system at GDT*

The gas dynamic trap (GDT) is a large-scale axially symmetric magnetic mirror device with 7-m-long central cell and two end-tanks, which house the expanding plasma flux. Design, diagnostics and physics of plasma confinement of this machine are described in Ref. 12, peculiarities related to ECRH are best described in Ref. 9. Particular parameters of the operation regime used for the present study are listed in Table 1.

As it was already noted, an increase of the electron temperature allows increasing the confinement time of NBI-born fast deuterons in the machine and, therefore, the neutron flux. For this purpose, GDT is equipped with ECRH system operating at the fundamental harmonic of the extraordinary wave [24]. The ECRH system consists of two 54.5 GHz, 450 kW gyrotron modules operating independently. Each module has its own waveguide transmission line and a beam launching antenna ("launcher"). The transmission line includes a two-mirror matching optics unit that converts gyrotron microwave beam into $HE_{11}$ mode at the entrance of a corrugated waveguide, a corrugated waveguide with inner diameter 63.5 mm, and three 90º miter bend units. One of these miter bends is a three-mirror universal polarizer which provides microwave beam polarization optimal for launching and absorption in the plasma. The launcher is a three mirror quasi-optical device, which allows adjusting the incident angle in two directions thus shifting the microwave beam intersection point with plasma both along and across the magnetic field lines. The launcher forms a Gaussian beam focused on the machine axis with the waist radius of 2.1 cm.

In the present experiment, we substitute one of the heating gyrotrons with a dedicated ECE receiver described in the next section. With one gyrotron, the actual (delivered to plasma) ECRH power is limited to 400±20 kW. The whole assembly is shown in Fig. 1. ECE measurements use all the advantages of the dedicated ECRH transmission system, such as good coupling efficiency, low transmission losses (about 10%) and the ability to control polarization using the universal polarizer. The angular pattern of ECE antenna naturally matches the pattern of the launcher. So, speaking in terms of geometric optics, the received ECE radiation propagates along the same rays as the heating wave beam, but in reverse direction. A polarization of ECE and ECRH lines can be controlled independently – we use both ordinary (O) and extraordinary (X) wave polarization for detecting ECE, and always use the X polarization for heating.

**Table 1.** GDT parameters used in ECE detection experiments

| | |
|---|---|
| Mirror-to-mirror distance | 7 m |
| Plasma radius at midplane | 0.2 m |
| Magnetic field at midplane | 0.32 T |
| Mirror ratio | 32 |
| ECRH power | 0.4 MW |
| ECRH frequency | 54.5 GHz |
| Electron temperature | 150-400 eV |
| Total neutral beam (NB) power | 4.5-5 MW |
| Absorbed NB power | 1.5-2 MW |
| NB pulse duration | 4.5 ms |
| NB injection angle | 45º |
| Energy of NB particles | 23 keV |
| Mean energy of fast ions | ~10 keV |
| On-axis plasma density at midplane | $1.5\text{-}2\times10^{19}\,\text{m}^{-3}$ |

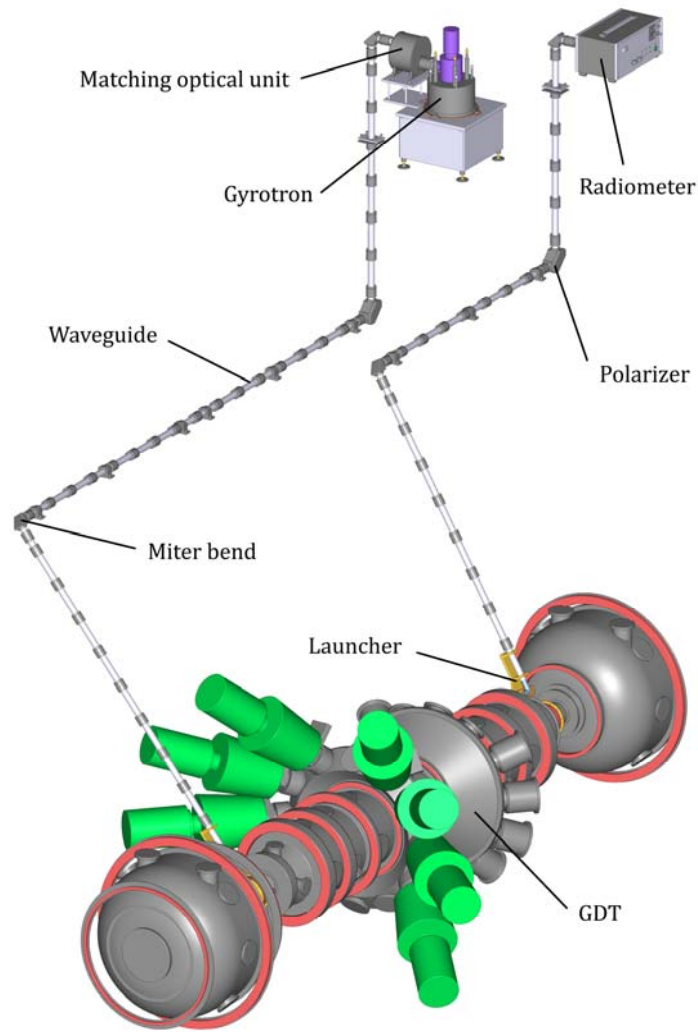

**Figure 1.** Assembly for ECE experiments

*B. ECE hardware*

Detection of ECE signal is performed with a dedicated heterodyne radiometer [23]. Operating frequency of the receiver is tunable in 53.5 – 59.5 GHz range with bandwidth 100 MHz, other parameters are listed in Table 2. Figure 2 shows a block scheme of the receiver. The input RF signal passes through the electronic p-i-n attenuator, which can either operate in an adjustable attenuation range of 0-40 dB, or in an external trigger mode with maximum attenuation of 50 dB. The last mode is used to protect the radiometer circuits against powerful gyrotron radiation. As such protection is critical during the experiments, we use an additional notch filter (>35 dB in 40 MHz band tuned to black out the gyrotron signal) and a high-pass filter with the cutoff frequency near the gyrotron frequency (30 dB at 54.5 GHz). Both components are installed inside the radiometer frame. A single side band balanced mixer is used to convert the received RF signal down to an intermediate frequency (IF). A local oscillator (LO) is based on Gunn diode with electronically tunable frequency (by a varicap diode) from 51.9 GHz to 58 GHz. LO is separated from the rest of the tract by a ferrite isolator. IF signal is increased by about 60 dB by a broadband IF amplifier. The frequency band of the radiometer is determined by 100 MHz bandpass filter with center frequency $1.63 \pm 0.01$ GHz. IF signal is detected by a square-law detector and comes to a two-stage video amplifier with 15+10 dB gain and 0-35 kHz band. Variable attenuator 0-25 dB is set between these components. The signal is recorded with an external ADC. The radiometer has an auxiliary calibration output, which is equipped with an integrating chain ($\tau = 1$ s). Calibration is performed using a black body microwave load with known radiation temperature. In this way, the ECE system is made ready for absolute measurements of the radiation temperature of the detected emission.

**Table 2.** Specifications of ECE radiometer.

| | |
|---|---|
| Operating frequency range | 53.5 – 59.5 GHz |
| HF bandwidth | 100 MHz |
| VF bandwidth | 35 kHz |
| Noise temperature | 0.2 eV |
| Fluctuation sensitivity | $3 \cdot 10^{-3}$ eV |

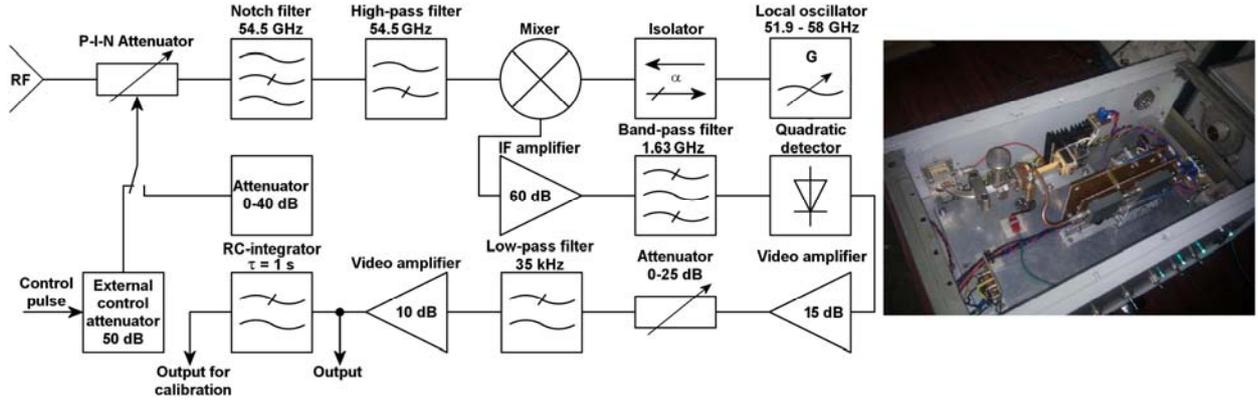

**Figure 2.** Block diagram and photograph of the radiometer.

*C. Specifics of ECRH/ECE experiment*

Figure 3 shows a detailed view of the launcher and an example of ray trajectories corresponding to the ECRH scheme. A microwave beam is launched obliquely into plasma near the magnetic mirrors from the high-field side at angle about 36° to the trap axis. Along its way to the first harmonic EC resonance, the beam experiences strong refraction, which is determined by a particular magnetic field distribution and a plasma density profile. As microwave beam propagates towards the ECR layer, a magnitude of the magnetic field decreases what eventually leads to internal reflection of some part the beam from the plasma-vacuum boundary. The plasma column acts as a waveguide, inhomogeneous in transverse and longitudinal directions, whereby the radiation is delivered to the ECR surface [19-21]. As we detect ECE through the same ECRH port, Figure 3 illustrates not only the ECRH geometry, but also the received emission pattern and the plasma configuration near the ECE antenna.

An important feature of the X wave propagation near the fundamental cyclotron harmonic is inaccessibility of the ECR layer (6 in Fig. 3) from the low magnetic field side due to presence of the evanescent region just behind the ECR (7 in Fig. 3). Exactly for this reason, the beam is launched from the high field side in the present ECRH scheme (4 in Fig. 3). However, since the gyrotron is located far away from the GDT in the low field, its radiation inevitably meets an auxiliary ECR surface from the low field side. In GDT conditions, this surface is located in the launching port inside the vacuum chamber, where some residual plasma may result in radiation reflection. This, so-called "parasitic", EC resonance plays an important role in optimization of ECRH, as well as in interpretation of ECE data. In particular, the parasitic ECR can be shifted from the plasma boundary to the vacuum chamber wall by increasing the current through the nearest axial coil of the GDT magnetic system (2 in Fig. 3), or by exploiting a small correction coil which is attached to the vacuum window (3 in Fig. 3). Using the auxiliary correction coil is preferable since it does not disturb the magnetic configuration in the main body and it is less demanding to power supply compared to the large coils. Figure 4 shows the effect of the correction coil on the parasitic resonance in more detail.

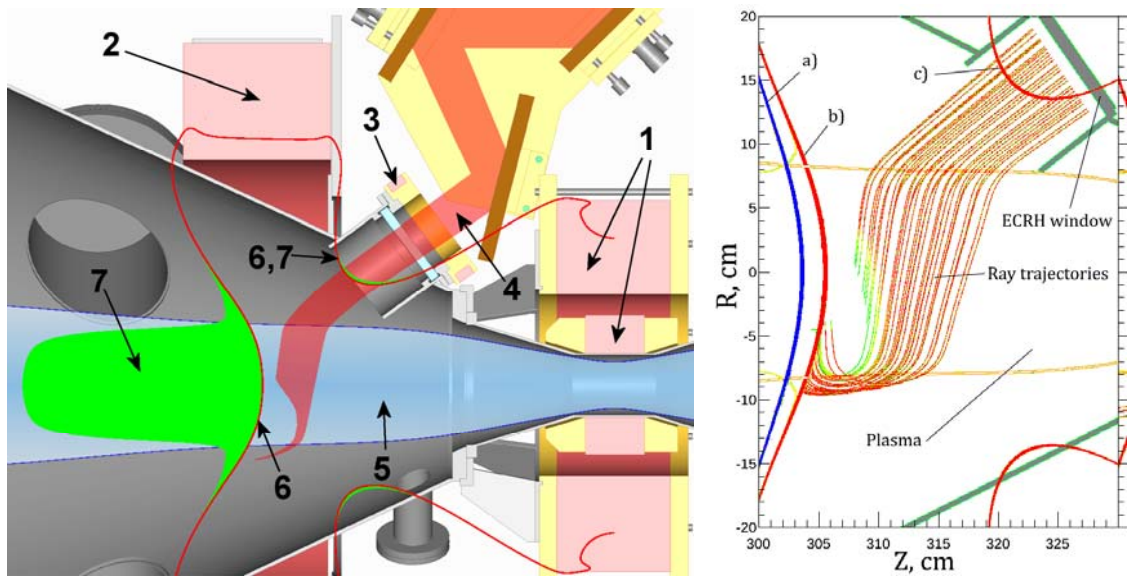

**Figure 3.** Left panel: ECRH/ECE launching conditions (1 - magnetic plug coils, 2 - cone coil, 3 - correction coil, 4 - microwave beam, 5 - plasma column, 6 - ECR surface, 7 - evanescent region in which the X mode cannot propagate). Right panel: typical ray trajectories in ECRH experiment (see Ref. 9).

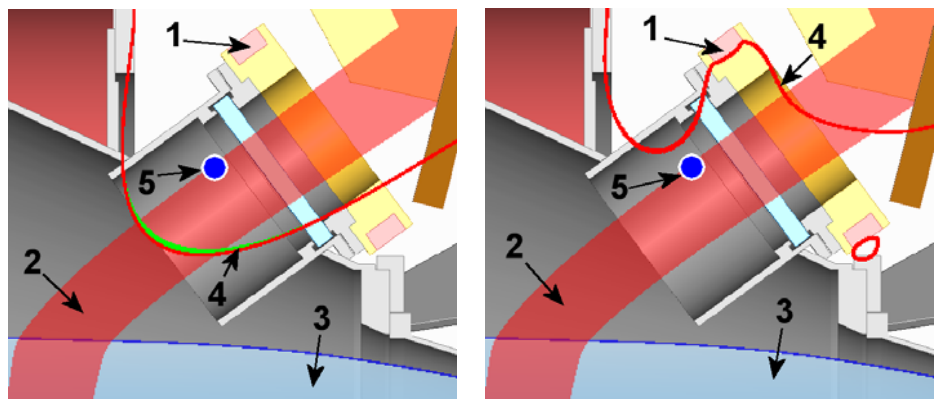

**Figure 4.** Parasitic resonance in ECRH/ECE launching port without (left) and with (right) the correction magnetic field applied (1 - correction coil, 2 - microwave beam, 3 - plasma column, 4 - ECR surface and region of evanescent X waves, 5 - reference point used in Figs. 5-7).

The experiment is conducted with deuterium plasma. A standard discharge scenario at GDT is as follows. Central cell vessel is pre-filled with neutral $D_2$ through a 10 ms gas feed pulse. The arc plasma gun located at one of the expander tanks is switched on from 0.5 ms to 3.6 ms. The gun fills the central part of the trap with a cold dense plasma. A potential of about 200 V is supplied to limiters at 4 ms. Plasma biasing results in the differential rotation of the plasma that effectively mitigates the development of MHD instabilities [25]. 5 MW NBI operates from 3.6 ms to 8.2 ms and results in accumulation of fast ions and plasma heating. In most shots with additional 400 kW ECRH, the gyrotron operates from 6.2 ms to 6.6 ms (or 7.1 ms). ECE level is measured at a fixed frequency band (one channel) during the entire discharge except for the time slots where the gyrotron is operating in shots with ECRH.

In experiments, we detected two types of ECE signal. In purely NBI discharges and at early stages of discharges with additional ECRH, the observed ECE was interpreted as a thermal emission of bulk electrons with Maxwellian distribution function. Contrary, at late stages of ECRH discharges, this thermal ECE was suppressed by a strong signal which may be attributed to the cyclotron emission of suprathermal electrons generated during ECRH.

### III. CYCLOTRON EMISSION OF THERMAL ELECTRONS

The first goal of the reported experiments was the validation of the current physical conceptions about the microwave heating of bulk electrons in the machine, in particular, the significance of the parasitic ECR surface in the launching port that might affect the X wave propagation in core plasma. Measured thermal ECE signals are the first step towards such validation.

*A. Experimental results*

Figure 5 (left panel) shows a typical evolution of measured X mode ECE in a discharge sustained only by 5 MW NBI in the standard GDT magnetic configuration. The level of ECE spectral intensity $I_\omega$ is characterized by the equivalent blackbody radiation temperature, $T_{\text{ECE}} = 2\pi\lambda^2 I_\omega$ with $\lambda$ being the wavelength of the received radiation, which we will further refer to as the effective ECE temperature. Measurements are carried out in the frequency range of 55.2 ± 0.05 GHz. There are two ECE signals measured with ($I_{\text{cc}}$ = 1 kA) and without ($I_{\text{cc}}$ = 0 kA) the correction coil used for elimination of the parasitic resonance in the launching port. The figure also indicates the evolution of the magnetic field in the launching port measured by a separate diagnostic coil. One can see that the difference between two ECE signals increases dramatically when the magnetic field exceeds a resonant value corresponding to the receiving frequency of 55.2 GHz. Switching on of the correction coil results in a further increase of the ECE level by a factor of two. Since the correction coil does not affect the main ECR, we can conclude that the parasitic resonance significantly affects the propagation of microwave radiation in the launching port (both for ECE and ECRH).

The bulk electron temperature $T_{\text{TS}}$ is measured in the central cross-section of the trap by Thomson scattering laser system, which yields one value per shot. The measured $T_{\text{TS}}$ and the time of this measurement are indicated in the left panel of Fig. 5. Note that the electron temperature measured by Thomson scattering is by about an order of magnitude greater than the effective ECE temperature: $T_{\text{TS}}$ = 180 eV versus $T_{\text{ECE}}$ = 22 eV. This difference suggests a potentially poor efficiency of ECRH in the studied configuration, at least at initial stages of ECRH. This is confirmed by ray-tracing simulations described below. At the same time, ECE measurements show that mitigation of the parasitic resonance by the correction coil may substantially improve the efficiency of ECRH. Indeed, in later experiments, we observed the improvement of the ECR-assisted plasma start-up [26, 27] with switching-on of the correction coil.

To emphasize the influence of the parasitic resonance we have performed a set of measurements with non-optimal magnetic configuration. To do this, we decreased by 6% the current an axial magnetic coil nearest to the main ECR zone (cone coil 2 in Fig. 3). With this modification both resonances shift in space: the main ECR moves about 5 cm to the magnetic mirror, and the parasitic ECR moves closer to the plasma boundary. So the density of residual plasma in a vicinity of the parasitic resonance is expected to be higher than in the standard configuration. The results of ECE measurements are presented in Fig. 6. Switching on of the correction coil leads to an essentially more pronounced effect. The ECE level increases by a factor of five; however, the effect starts later and lasts shorter due to a decreased magnetic field in the main volume of the trap. Note again a big difference between the bulk electron temperature measured with TS diagnostics and the ECE temperature ($T_{TS}$ = 150 eV versus $T_{ECE}$ = 10 eV).

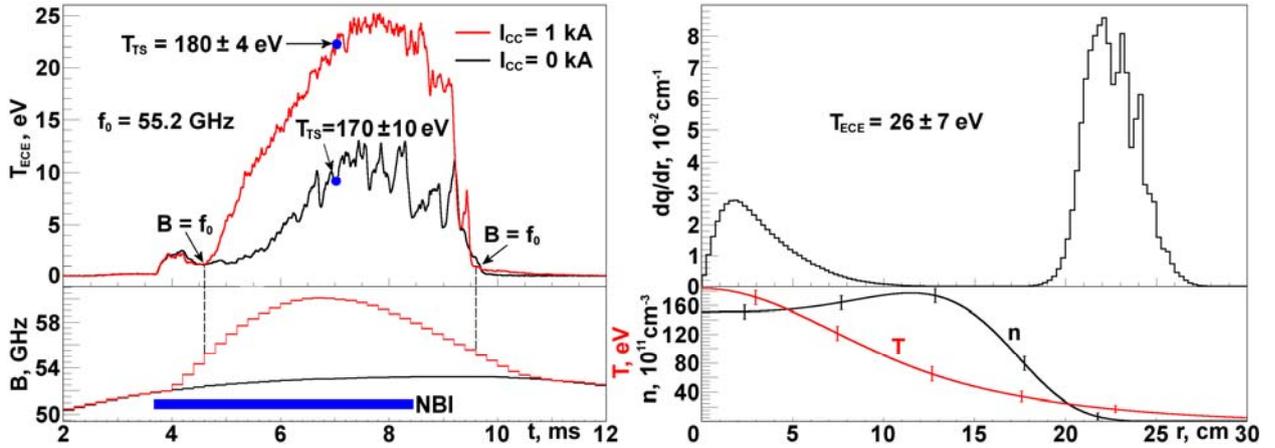

**Figure 5.** Left panel: measured ECE temperature (X mode at 55.2 ± 0.05 GHz) and magnetic field evolution in the pure NBI discharge in the standard magnetic configuration with (red lines) and without (black lines) the correction field for the elimination of the parasitic resonance (adopted from Ref. 23). The magnetic field is shown in units of corresponding electron cyclotron frequency and is measured at the point inside the ECRH launching port marked by label 5 in Fig. 4. Right panel: radial distributions of the specific emissivity (top) provided by the ray tracing calculations for the electron density and temperature profiles (bottom) corresponding to the experimental conditions shown in the left panel. The density and temperature are measured by Thomson scattering diagnostics in the central plane; all profiles are related to the central plane.

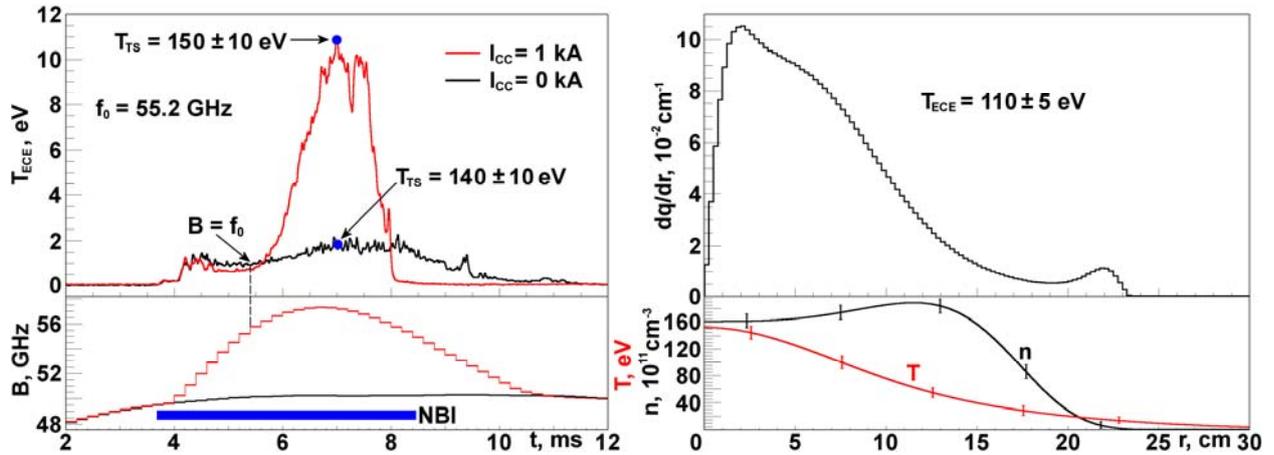

**Figure 6.** Same as Fig. 5 but for reduced magnetic field in the trap body under ECRH port.

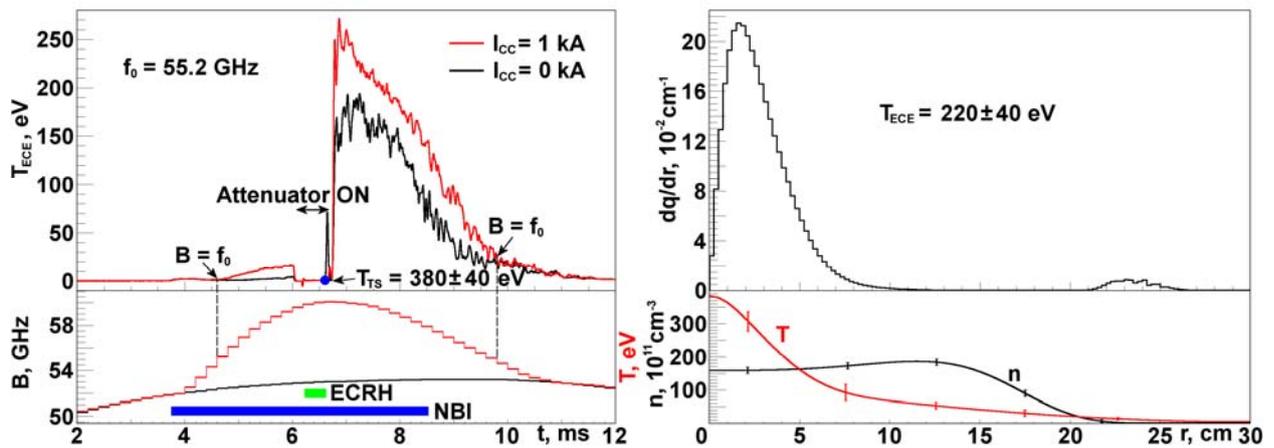

**Figure 7.** Same as Fig. 5 but for combined 5 MW NBI / 400 kW ECRH discharge in the standard magnetic configuration.

The difference between the electron temperature and effective ECE temperature becomes essentially less pronounced in discharges with additional 400 kW ECRH applied for a short period (0.4 ms) during the 5 MW NBI heating, see Fig. 7. For the protection of the radiometer, the ECE receiver is blinded by a controlled 40 dB attenuator during the gyrotron operation. Comparison with similar shots without ECRH shows that the additional heating leads to increase of the electron temperature by about a factor of two (from $T_{TS}$ = 180 eV to 350 eV), while the ECE level rises by order of magnitude (from $T_{ECE}$ = 25 eV to 250 eV). Close values of the radiation and kinetic temperatures, $T_{ECE} \approx T_{TS}$, indicate that the plasma column becomes optically thick for the radiation with the electron temperature increase. Thus, ECRH in the GDT conditions adjusts plasma parameters in such a way that the absorption of microwaves by plasma is itself improved. In ECRH shots the effect of the correction coil is also present, but it is significantly less pronounced.

We performed a few shots in attempt to measure ECE in the orthogonal (O mode) polarization. Due to large experimental uncertainties, we do not present the O mode data, but some preliminary conclusions may be formulated. Thermal ECE intensity in the O mode (in pure-NBI discharges) is generally by 1-2 orders of magnitude lower than the level of the X mode emission. It is not affected by the correction coil in agreement with theoretical predictions since there is no evanescent region for the O mode waves in the GDT conditions. On the other hand, in ECRH discharges the O mode signal is strongly affected by the emission of fast electrons during the whole discharge (as opposite to the X mode). We discuss this issue in Sec. IV.A.

*B. Discussion*

Significant attenuation of ECE by the parasitic resonance may be viewed as tunneling of the X mode propagating in an inhomogeneous magnetic field through the evanescent region near the upper-hybrid resonance, which in our conditions coincides with the fundamental harmonic electron cyclotron resonance, see Fig. 8 (left). This effect is described by the classical formula for the tunneling efficiency,

$$T = \exp(-\pi \omega_p^2 L_B / c\omega),$$

where $T$ is a ratio between X wave intensities before and after the evanescent layer, $\omega_p = (4\pi e^2 n_e / m_e)^{1/2}$ is the electron Langmuir frequency (here and below we use the Gaussian system), $\omega$ is the angular wave frequency which is close to the electron cyclotron frequency $\omega_B = eB/m_e c$, and $L_B = B/(dB/dl)$ is the scale of magnetic field variation in the direction of wave propagation. This result may be obtained as an exact solution of wave equation corresponding to cold plasma with constant density in a linearly varying magnetic field, see Ch. 5 § 27 in Ginzburg [28] or § 6.2.3 in Swanson [29]. In the right panel of Fig. 8, we plot the tunneling efficiency as a function of the residual plasma density for the actual GDT parameters. On the other hand, this efficiency may be estimated directly from our ECE measurements as a ratio of radiation levels before and after correction coil eliminates the parasitic resonance in the launching port, $T = T_{ECE}^{c.c.off} / T_{ECE}^{c.c.on}$. In this way, experimental data described in the previous may be projected to the right plot in Fig. 8.

One can find that the density of residual plasma varies between $5 \times 10^9$ and $3 \times 10^{10} \, cm^{-3}$. The lower limit corresponding to shots with ECRH is somewhat unexpected: it seems that ECRH somehow cleans out the residual plasma in the opposite launching port. The upper limit corresponds to shots with a reduced magnetic field when the parasitic resonance shifts closer to the main plasma.

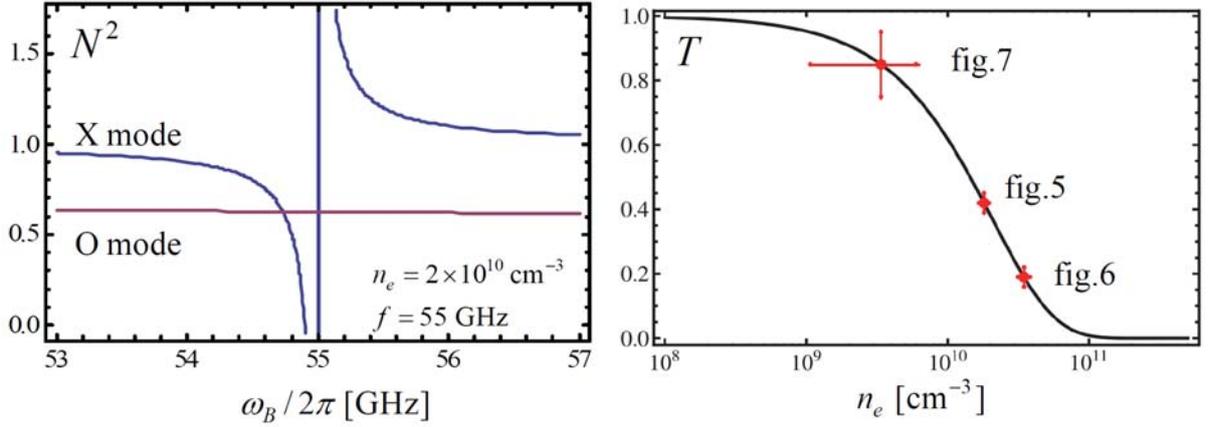

**Figure 8.** Left panel: refractive index for the X and O modes as a function of the cyclotron frequency near the parasitic resonance in the launching port, evanescent region corresponds to $N^2 < 0$. Right panel: transmission efficiency through the evanescent layer as a function of the residual plasma density. Points with error bars indicate the values corresponding to the ECE measurements shown in Figs. 5 – 7. GDT parameters are $\omega/2\pi = 54.5\,\text{GHz}$ and $L_B = 50\,\text{cm}$.

For simulation of ECE of inhomogeneous plasma in GDT, we use the ray-tracing code previously developed for the ECRH modeling at this device [9]. The microwave power coming into the ECE antenna is modeled as a sum over a set of rays distributed according to the antenna pattern. The intensity $I_\omega$ of the emitted radiation along each ray can be calculated from the same radiation transfer equation as used for the ECRH modeling but with spontaneous emissivity taken into account [22, 30],

$$N_r^2 \frac{d}{dl}\left(\frac{I_\omega}{N_r^2}\right) = j_\omega - \alpha_\omega I_\omega, \quad j_\omega = \frac{N_r^2 \omega^2 T_e}{(2\pi)^3 c^2} \alpha_\omega.$$

Here $\alpha_\omega$ is the absorption coefficient calculated for warm magnetized plasma with the Maxwellian distribution function of electrons in non-relativistic approximation [30], $j_\omega$ is the thermal emissivity defined by Kirchhoff's law in Rayleigh–Jeans limit [22], $N_r$ is so called ray refraction index for a selected electromagnetic mode [22], and $T_e$ is the distribution of the electron temperature along the ray trajectory $l$. For spontaneous emission of equilibrium plasma, the radiation transfer equation may be expressed in a simpler form if the intensity of the emitted radiation is measured in terms of effective radiation temperature,

$$\frac{dT_r}{dl} = \alpha_\omega (T_e - T_r), \quad T_r = \frac{(2\pi)^3 c^2}{N_r^2 \omega^2} I_\omega.$$

Spontaneous emission may be described by the solution of this equation corresponding to semi-infinite ray $l \in (0, \infty)$ (that starts at the antenna at $l = 0$ and propagates into plasma) with boundary condition $T_r(\infty) = 0$,

$$T_r(0) = \int_0^\infty T_e(l) \left\{ \alpha_\omega(l) \exp\left(-\int_0^l \alpha_\omega(l')dl'\right) \right\} dl .$$

The measured ECE signal is a sum of such solutions over all rays weighted according to the angular pattern of the receiving antenna, $T_{ECE} = \sum w_i T_{ri}(0)$ where index $i$ denotes different rays. This quantity may be calculated numerically using only the solution of the ECRH problem. Indeed, the term in curly braces is the specific power absorbed per unit element along the ray that is found during ECRH modeling, i.e. for $j_\omega = 0$ and $I_\omega(0) = I_0$,

$$\frac{dq}{dl} \equiv \frac{\alpha_\omega I_\omega}{N_r^2 I_0} = \alpha_\omega \exp\left(-\int_0^l \alpha_\omega(l')dl'\right) .$$

Therefore, the net ECE temperature is

$$T_{ECE} = \sum w_i \int T_e \, dq_i = \int_0^{r_{max}} T_e(r) \frac{dq}{dr} dr ,$$

where

$$\frac{dq}{dr} = \sum w_i \int_0^\infty \frac{dq_i}{dl} \delta(r - r_i(l)) dl$$

is the ECRH power deposition profile over magnetic flux surfaces labeled with the radial coordinate $r$ at the central cross-section of the trap. In this way, any ECRH code that calculates radial power deposition profile may be easily adopted for calculation of ECE of thermal plasma. In this paper, the power deposition is provided by the available ECRH ray-tracing code, which has been recently benchmarked against the more accurate quasi-optical code $QOOT$ [21].

The result of calculations of ECE intensity is indicated in the right panels in Figures 5 – 7. Note that the influence of the evanescent layer near the parasitic ECR in the launching port is neglected in this approach. The power deposition profile allows visualizing the origin of a received signal and thus characterizes the locality of a potential diagnostics of the electron temperature based on ECE measurements.

One can see that measured and calculated $T_{ECE}$ are in agreement within error bars for both types of discharges in the standard magnetic configuration: for optically thin plasma of pure NBI discharge (Fig. 5) and optically thick plasma of ECRH discharge (Fig. 7) provided that the parasitic resonance is eliminated with the correction coil. However, ray-tracing essentially overestimates the ECE level for the reduced magnetic field (110 eV against measured 10 eV, Fig. 6).

The transition from optically thin to thick plasma with electron temperature increase may be understood qualitatively using the following formula for the emissivity in the X mode at the fundamental harmonic:

$$\alpha_\omega \approx K\sqrt{T_e}\exp\left(-\frac{W}{T_e}\right), \quad W = \frac{(\omega-\omega_B)^2}{2k_\parallel^2}m_e c^2,$$

where $K$ is a constant independent of $T_e$, and $W$ is the energy corresponded to the Doppler shift in a cyclotron resonance condition [30]. In the GDT conditions $W \approx 625$ eV. From the experimental data, we know the optical depth $\eta = \int_0^\infty \alpha_\omega(l)dl \approx T_{ECE}/T_{TS} \approx 0.15$ at $T_e \approx 180$ eV (see Fig. 5). From that, we can reconstruct the dependence of optical depth over the electron temperature as $\eta \approx 0.36\sqrt{T_e[\text{eV}]}\exp(-625\,\text{eV}/T_e)$. One can find that this is a rapidly increasing function of $T_e$ that reaches unity at $T_e \approx 330$ eV.

## IV. CYCLOTRON EMISSION OF SUPRATHERMAL ELECTRONS

A strong ECE signals were repeatedly observed at the late stage of the discharge in the experiments with additional ECRH. These signals are detected in both (X and O) plasma polarizations after all the plasma heating is off (NBI and ECRH) and plasma starts to decay. What is more intriguing, the signals are also observed even after the total decay of the main plasma in the trap volume. These features allow interpreting these signals as ECE from a population of fast electrons produced by strong ECRH. With energies of 100 keV and higher, these electrons are confined adiabatically. They interact weakly with the bulk plasma and can live in the trap much longer than the main discharge duration.

*A. Experimental results*

A typical example of suprathermal ECE data is presented in Fig. 9 which shows some ECE signals for various frequencies and polarizations, the magnetic field near the axis (in frequency units) measured by a probe coil in a calibration shot, and the line-averaged plasma density. Each ECE signal is measured in a separate discharge of the series of identical shots. Note that the radiometer input was blocked during the gyrotron operation. One can see that the X mode ECE level substantially increases during the ECRH, then it decreases as the plasma decays, but then it sharply increases again when the bulk plasma has decayed completely. The O mode signal does not have a gap during the plasma decay. Instead, it monotonically decreases still maintaining a relatively high level at the late stage with no bulk plasma. Note also a strong emission in the O mode during the developed stage of the discharge when NBI-heating is on. This indicates that the O mode emission of fast electrons is not screened by the bulk plasma, and it dominates during the main discharge. At the same time, a good agreement between measured X mode ECE and the thermal level predicted by the ray-tracing calculations is an evidence of low contribution of fast electrons to the X mode emission.

Figure 9 also shows the signal of the standard neutron detector, which may also measure the X-rays during the reported experiment. Indeed, the neutron flux stops after approximately 1 ms after the end of the NBI. Therefore, the rest of the signal may be attributed only to the X-ray radiation produced by fast electrons with energies of at least 100 keV (as required for the transmission of X-rays through 5 mm thick stainless steel wall of the vacuum chamber). Similar correlated signals are also observed in potential oscillations of the plasma electrodes at the trap ends.

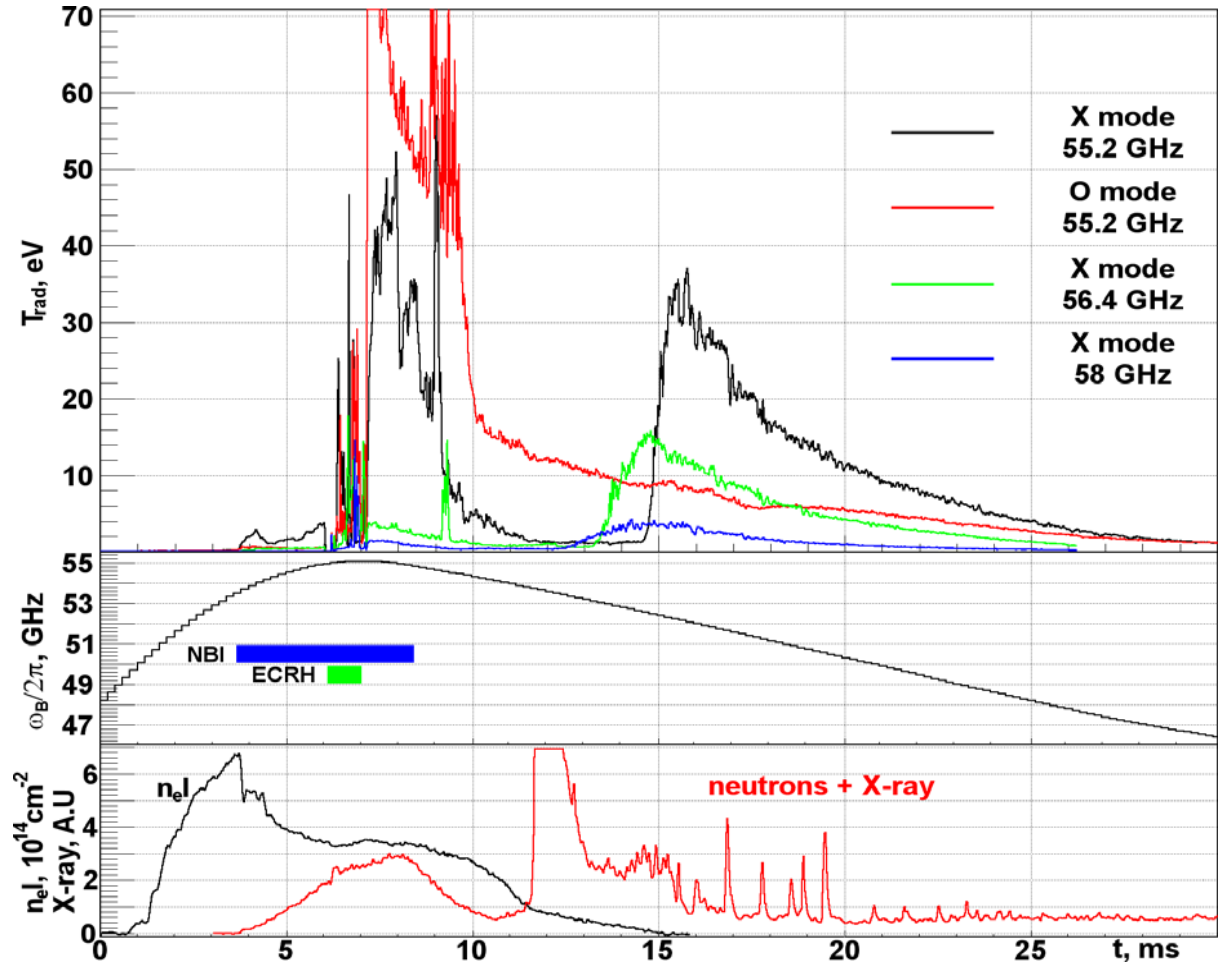

**Figure 9.** ECE temperature at different frequencies and polarization modes, electron cyclotron frequency at a fixed point on the GDT axis, X-ray signal and line-averaged plasma density in a discharge with combined 5 MW NBI / 400 kW ECR heating in the standard magnetic configuration. Adopted from Ref. 23.

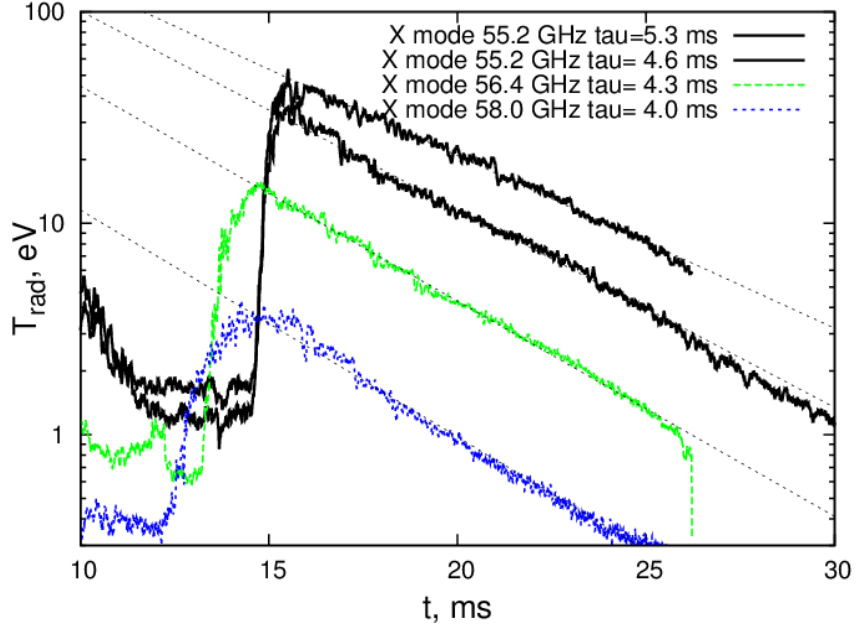

**Figure 10.** ECE temperature from Fig. 9 in logarithmic scale. Signals are approximated by the exponential function $a\exp(-t/\tau)$ with characteristic times $\tau$ indicated in the legend.

Figure 10 shows the same ECE signals as in Fig. 9 but in a logarithmic scale. One can see that ECE intensity falls according to the exponential law $\exp(-t/\tau)$ with practically the same time scale $\tau \approx 4-5$ ms for all shots. This is a common feature of all measured ECE signals at the late stage of plasma decay (totally, we have about a hundred successful shots).

*B. Discussion*

Fast electrons can be easily generated by strong ECRH. In GDT conditions their fraction is moderate compared to other devices – indeed, an effective power deposition into the thermal plasma component is considered as one of the major achievements of the GDT ECRH experiment [10]. Nevertheless, suprathermal electrons play an essential role in the ECR plasma start-up recently implemented at GDT [26, 27]. According to our present understanding, the fast electrons have energies $\sim 10$ keV and density $\sim 10^{10}$ cm$^{-3}$; they are entirely responsible for gas ionization and plasma pressure at the initial stage of gas breakdown and the subsequent stage of seed plasma build-up [27, 31]. Generation mechanism of such electrons is discussed in detail in Ref. 27, where a possibility for further acceleration of a minor fraction of electrons up to relativistic energies is not excluded.

Assuming that rarefied plasma at the late phase of discharge decay is transparent, one can estimate the intensity of ECE that is coming to the antenna as

$$I_\omega = \int dl \int w(\mathbf{k}) d^2\mathbf{k} \int P_\mathbf{k}(\mathbf{v}) f(\mathbf{v},l) d^3\mathbf{v},$$

where $l$ is the viewing $l$ the coordinate along the antenna beam, $w(\mathbf{k})$ characterizes the angular antenna pattern with the norm $\int w(\mathbf{k})d^2\mathbf{k} = 1$, $f(\mathbf{v})$ is the distribution function of fast electrons, and $P_\mathbf{k}(\mathbf{v})$ is the power radiated by a single electron spinning in a magnetic field [22],

$$P_\mathbf{k} = \frac{e^2 \omega_B^2 (1+\cos^2\theta)}{4\pi c^3} \upsilon_\perp^2 \, \delta(\omega - k_\parallel \upsilon_\parallel - \omega_B/\gamma).$$

The argument of the delta-function represents the standard cyclotron resonance condition with radiation frequency $\omega$, cyclotron frequency $\omega_B = \omega_B(l)$, electron velocities $\upsilon_\parallel$ and $\upsilon_\perp$ along and across the magnetic field, longitudinal wave vector $k_\parallel = \omega \cos\theta / c$, and the relativistic factor $\gamma$ of the radiating electron. Here we take into account only the fundamental harmonic and neglect others assuming weakly relativistic electrons.

Performing integration over $l$, one can obtain the estimate for the effective ECE temperature,

$$T_{\text{ECE}} = 2\pi r_e \lambda L_{\text{ECE}} \frac{\omega_B^2}{\omega^2} \frac{1+\cos^2\theta}{\cos\theta} w_\perp, \qquad w_\perp = \tfrac{1}{2} <\int m\upsilon_\perp^2 f(\mathbf{v}, l_{\text{res}})d^3\mathbf{v}>.$$

Here $r_e = e^2/mc^2 \approx 2.8 \cdot 10^{-13}$ cm is the classical electron radius, $\lambda \approx 0.5$ cm is the received wavelength, $l_{\text{res}}(\upsilon_\parallel, \upsilon_\perp)$ is the coordinate of the resonance, $L_{\text{ECE}} \equiv c\, \partial l_{\text{res}}/\partial \upsilon_\parallel \approx L_B$ where $L_B = B/(\partial B/\partial z) \approx 50$ cm is the scale of magnetic field variation along the trap axis, and $w_\perp$ is the density of energy stored in gyromotion of emitting resonant electrons averaged over the antenna pattern. Taking $T_{\text{ECE}} \sim 50$ eV and a reference energy of the emitting electrons of 10 keV, one obtains that their averaged density over the viewing volume is about $5 \cdot 10^7 \text{cm}^{-3}$. This is by more than two orders of magnitude less than is expected for the ECR plasma breakdown [27]. Thus, either only a minor fraction of hot electrons can radiate at the frequencies and angles received by our antenna, or hot electrons are suppressed during the developed stage of NBI discharge.

Note that the received ECE signal is sensitive to a mean energy of electrons with particular velocities defined by the ECR condition. This condition varies essentially during the discharge decay due to variation in time of the confining magnetic field. For the example shown in Fig. 9, the cyclotron frequency shift is $\Delta\omega_B/\omega \approx 0.1$ in the interval between 15 and 25 ms, what corresponds to the equivalent Doppler shift, $\Delta\omega_B = k_\parallel \upsilon_\parallel$, by electrons with the longitudinal energy about 5 keV or to the relativistic shift, $\Delta\omega_B = (\gamma - 1)\omega$, by electrons with an energy of about 100 keV. These shifts are by one order of magnitude greater than the uncertainty introduced by finite antenna pattern. So our first intention was to explain the variation of the non-thermal ECE by its dependence on the distribution function of emitting electrons [23]. This assumption was supported by estimation of the confinement time governed by Coulomb collisions with the cold plasma, that appeared to be much longer that the ECE variation time. Indeed, assuming the background plasma density $n_e = 10^{11} \text{cm}^{-3}$ (as the lowest density measured by the dispersion interferometer) and mirror ratio $R = 30$, one obtains Pastukhov's confinement time of fast electrons with energy $E$ as

$$\tau_c = \nu_{ei}^{-1} \ln R \approx (E/10 \text{ keV})^{3/2} \text{ sec} > 1 \text{ sec}.$$

Therefore, reduction of the ECE level may be only correlated with the shift of the resonance condition to higher energies at which the number of particles is decreasing. This provides a key to reconstructing the distribution function of the emitting electrons.

However, more elaborate analysis convinces us that this natural idea is incompatible with our data. The first objection was due to the assumption that most of the emitting electrons are localized near the trap axis. However, study of the ECR breakdown at GDT suggests that fast particles have a broad and nearly flat radial distribution over the whole plasma volume [27]. In this case, the ECR condition for emitting at one frequency electrons broadens after averaging over the volume viewed by the antenna due to spatial inhomogeneity of the magnetic field. Spatial variation of the magnetic field becomes of the same order as its temporal variation; therefore the details of the velocity distribution function are not pronounced in the ECE data as clear as for the axially localized fast electrons.

Second and the main argument against the "resonant" interpretation of ECE data is that we can not explain strictly exponential dependencies well illustrated in Fig. 10. It seems unbelievable that fast electron tails with 100 keV range reproduce the exponent; moreover, this exponent remains the same from shot to shot. A more credible hypothesis is that the exponential dynamics is governed by a decay of the residual cold plasma, which occurs with an approximately constant time. As the simplest assumption, we may consider gas-dynamic losses in a zero-dimension (volume averaged) approximation,

$$\frac{dn_e}{dt} = -\frac{n_e}{\tau_{gd}}, \quad \tau_{gd} = \frac{RL}{2\upsilon_s} \approx \text{const}.$$

Here $\tau_{gd}$ is the gas-dynamic confinement time defined by the mirror ratio $R$, the trap length $L$, and the ion-acoustic velocity $\upsilon_s = \sqrt{(T_e + T_i)/M}$ in the background plasma [32]. Variation of the gas-dynamic time is small because it does not depend on the density, and the temperature settles on much faster time-scale than the density decay. To explain the exponents in Fig. 10, one must assume $\tau_{gd} = 5$ ms or $T_e + T_i \approx 10$ eV in GDT conditions, which is quite reasonable.

A trickier question is how the cold plasma density affects the ECE of hot electrons. This may be understood if we consider electron-cyclotron instabilities caused by hot electrons. The interaction of unstable waves with resonant electrons leads to the diffusion of energetic electrons in velocity space on a much faster than collisional time-scales. As a result, some fast electrons fall into the loss cone and leave the trap. These losses reduce the instability growth rate and, finally, restrict the increase in the electromagnetic energy density in the system. In Ref. 33 we consider such process in a particular case of decaying plasma confined in a mirror trap. The main conclusion that is relevant to the present study is the following. Nonlinear evolution of wave turbulence and fast particles results in the system oscillations around the instability threshold, $\gamma_L = \nu$. Here $\gamma_L$ is the linear growth rate of an unstable wave that depends, in most cases linearly, on the density of fast electrons, and $\nu$ is the damping rate of the same wave due to the presence of cold background plasma. In most cases $\nu$ is proportional to electron-ion collision rate in the background plasma. While the background plasma decays, the damping rate $\nu(t)$ reduces, so that $\gamma_L > \nu$. At that moment the instability develops, a portion of fast electrons is ejected from the trap, so that system goes again under the instability threshold $\gamma_L < \nu$ and the process repeats, see Fig. 11. As a result, the average density of hot electrons follows exactly the density of cold plasma to provide the near-threshold conditions during the entire time of the plasma decay. Previously this concept has been verified experimentally in an axisymmetric open trap in conditions very similar to GDT case: in a decaying plasma after the ECR discharge supported by a high-power gyrotron [34-37].

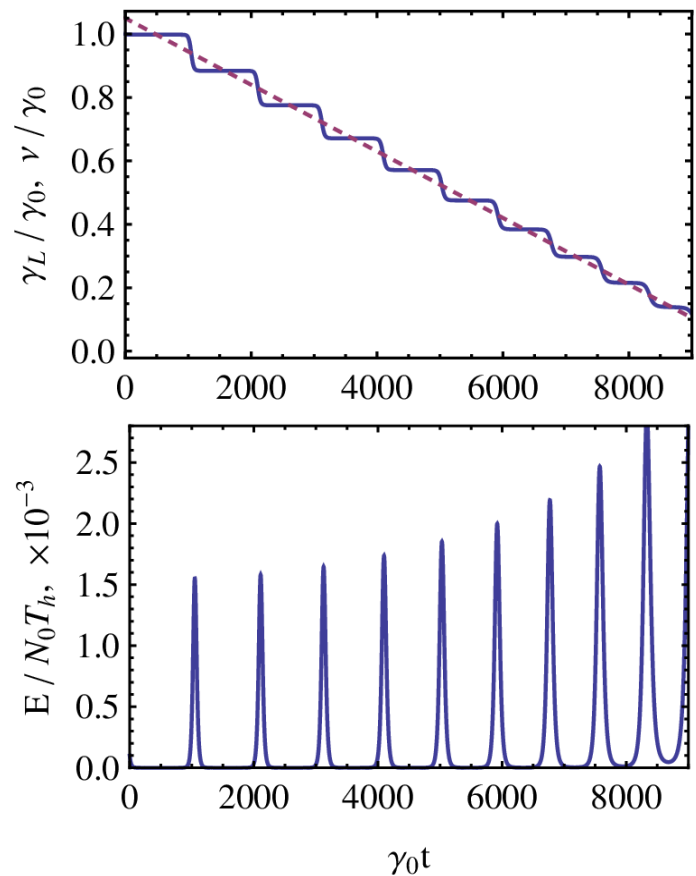

**Figure 11.** Time dependence of the growth rate $\gamma_L$ (top, solid line), the damping rate $\nu$ (top, dotted line), and the density of electromagnetic energy of an unstable mode (bottom) in a decaying plasma. For more details see Ref. 33.

Thus, our current interpretation of the ECE signals shown in Fig. 9 is that they more likely follow the total density of fast electrons rather than their velocity distribution function. The losses of fast electrons are dominated by plasma micro-instabilities that couple their density to the density of the residual cold plasma. Observed bursts of the X-ray radiation correlated with the fast drops of ECE signal (see Fig. 9) and variation of the end-plate potentials are evidence in favor of this concept. However, a more elaborate investigation aimed at validation of the cyclotron maser concept is being under preparation.

The last question to be addressed is the nature of the gap between the decay start and the suprathermal X mode emission, see time interval 10-15 ms in Fig. 9. This gap is likely related to the parasitic resonance in the ECRH port, which is not affected by the correction coil since at those times the coil is off. The moment of ECE switching-on correlates well with the bifurcation of time-varying ECR surfaces shown in Fig. 12. In the high magnetic field the ECR surface covers the antenna, but as the magnetic field falls, the ECR surface "jumps" away and opens the antenna mouth. At each frequency this happens with different time delays, and these delays fit exactly to the X mode ECE switching-on at various frequencies – the higher the frequency, the sooner it is triggered on. In the additional experimental campaign, we observe the same correlation for the X mode ECE below the gyrotron frequency in 53.5 – 54.4 GHz range.
.

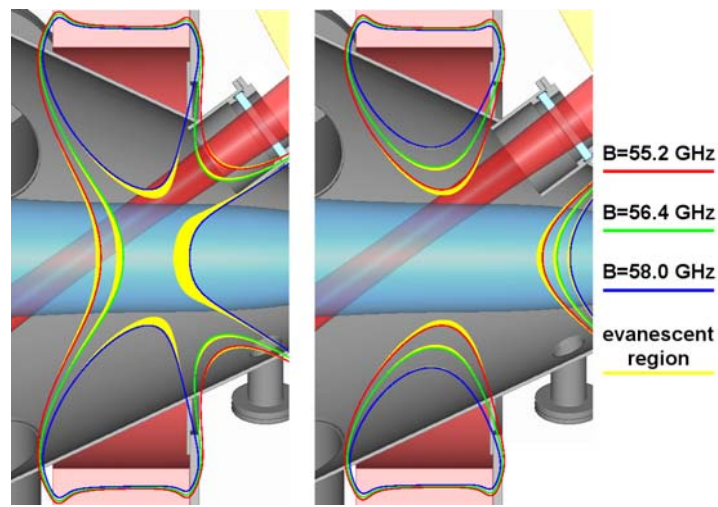

**Figure 12.** Bifurcation of ECR surfaces $\omega_B = \omega$ at different frequencies as a possible trigger of X-mode EC emission of fast electrons during the suppressed ECE stage (left panel), and after the ECE signal appears again (right panel).

## V. SUMMARY

The new ECE diagnostics has been installed to augment the ECRH experiment at the GDT device. The particularities of plasma cyclotron emission in the vicinity of the ECRH frequency were studied experimentally for different discharge scenarios. Measured thermal emission has partly validated the existing physical conceptions about the microwave plasma heating in the machine, in particular, the auxiliary ECR surface in the launching port was shown to affect the wave propagation towards the core plasma. The efficiency of an additional magnetic coil for local suppression of the auxiliary resonance has been proven by the thermal ECE measurements. Against our expectations, it has been found that ECRH may start from a weakly absorbing plasma, but during the discharge the evolution of the electron temperature leads to an eventual transition to the optically thick plasma. Before these findings we believed that ECRH of bulk plasma always starts with the optically thick plasma in GDT conditions.

Measured non-thermal ECE have unambiguously confirmed the existence of suprathermal electrons in GDT. These electrons are generated by the intense microwave field during the ECR heating of the main plasma. Explanation of time dependence of ECE level observed in the varying magnetic field and decaying plasma is a rather challenging task. Our current hypothesis is based on the concept of stimulated micro-instabilities that cause fast losses of suprathermal electrons, which may be registered via the detection of spontaneous electromagnetic emission. However, further experiments are needed to confirm this idea.

## ACKNOWLEDGEMENTS

The authors would like to thank F. F. Arkhiptsev and V. A. Genneberg for their invaluable support with the development of the radiometer, and the GDT Team for help during the experimental campaign. The research was supported by the Russian Science Foundation (pr. 14-12-01007).